\documentclass[12pt, epsfig]{article}
\usepackage{amsmath,amssymb,amsfonts,epsfig,graphicx}

\newcommand{\bse}{\begin{subequations}}
\newcommand{\ese}{\end{subequations}}
\newcommand{\be}{\begin{equation}}
\newcommand{\ee}{\end{equation}}
\newcommand{\bea}{\begin{eqnarray}}
\newcommand{\eea}{\end{eqnarray}}
\newcommand{\ba}{\begin{array}}
\newcommand{\ea}{\end{array}}

\newcommand{\bc}{\begin{center}}
\newcommand{\ec}{\end{center}}
\makeatletter \@addtoreset{equation}{section}

\makeatother
\begin{document}

\baselineskip 18pt%

\begin{titlepage}

\vspace*{1mm}%
\hfill%
\vbox{
    \halign{#\hfil        \cr
           hep-th/0611236 \cr
           IPM/P-2006/064 \cr
           } % end of \halign
      }  % end of \vbox
\vspace*{15mm}%

\bc {{\Large {\bf Stabilization of Compactification Volume\\
In a Noncommutative Mini-Super-Phase-Space
}}}%
\vspace*{0.5cm}

{\bf N. Khosravi$^1$, H. R. Sepangi$^{1}$, M. M. Sheikh-Jabbari$^2$}%
\vspace*{0.5cm}

{\it {$^1$ Department of Physics, Shahid Beheshti University,
Evin, Tehran 19839, Iran}}\\
{E-mail: {\tt n-khosravi, hr-sepangi@sbu.ac.ir}}%
\vspace*{0.2cm}

{\it {$^2$Institute for Studies in Theoretical Physics and Mathematics (IPM)\\
P.O.Box 19395-5531, Tehran, Iran}}\\
{E-mail: {\tt jabbari@theory.ipm.ac.ir}}%
\vspace*{1.5cm}
\end{center}

\begin{center}{\bf Abstract}\end{center}
\begin{quote}

We consider a class of generalized FRW type metrics in the context
of higher dimensional Einstein gravity  in which the extra
dimensions are allowed to have different scale factors. It is
shown that noncommutativity between the momenta conjugate to the
internal space scale factors controls the power-law behavior of
the scale factors in the extra dimensions, taming it to an
oscillatory behavior. Hence noncommutativity among the internal
momenta of the mini-super-\emph{phase}-space can be used to
explain stabilization of the compactification volume of the
internal space in a higher dimensional gravity theory.

\vspace{3mm}\noindent
\end{quote}%
\end{titlepage}

\newpage
\section{Introduction}
It is a long standing problem that gravity (the Einstein Hilbert
action) is plagued with severe non-renormalizability and string
theory has appeared as the only consistent theory of quantum
gravity with a reasonable low energy limit. Consistent string
theories are, however, all formulated in dimensions higher than
four, the best understood of them in ten dimensions. Besides the
extra dimensions consistency of string theory requires
supersymmetry. Neither the extra dimensions nor supersymmetry are
not backed by the current observational data. Hence, to turn
string theory to a viable physical model one needs to address the
above issues. In this work we will concentrate on the extra
dimensions and leave the supersymmetry to other works.

Theories in higher than four space-times dimensions was considered
by Kaluza-Klein (KK) in the early days of Einstein gravity, with
the hope of unifying gauge theories with gravity. In the KK setup
the question ``why the extra dimensions are not observed today?''
is answered by considering the extra dimensions to be wrapping a
compact manifold of a small (Planckian) size. This is essentially
what is also used in modern day string theory.  Namely, the six
extra dimensions are compactified on a Ricci flat complex,
K\"ahler three-fold, a Calabi-Yau three-fold ($CY_3$) whose volume
is of the size of strings (or Planck) scale
\cite{String-books}.\footnote{There is also the Randall-Sundrum
(RS) scenario (which is also called RS ``compactification'')
\cite{RS} which is not based on the idea of compact Ricci flat
manifold as extra dimensions, but employs the warped metric to
make the extra dimensions invisible to the ``low energy'' physics
observed today.} Using (CY) compactification idea one immediately
faces the ``compactification  and moduli problem'', i.e. { why and
how among so many possibilities for the shape and size of the
$CY_3$ a specific one, which has resulted in the Universe and
physics we see today, has been chosen?} In other words, the
question is whether string theory bears a (dynamical) mechanism
which lands us in a specific point in the so-called string
landscape \cite{Lenny}.

Recent developments in string theory compactification when various
form-field fluxes are turned on, the flux compactification, has
taught us that the shape moduli (complex structure moduli) can be
fixed once the fluxes are chosen \cite{GKP}. The size moduli
(K\"ahler structure moduli) and in particular the overall volume
modulus are, however, more involved and their fixing is somewhat
less  established and clear in the flux compactification setup,
e.g. see \cite{MN, Keshav, Shamit} in the string theory context
and \cite{Nima, Zhuk, Carroll} in the context of large extra
dimension models.

It is, of course,  desirable to invoke a more dynamical reasoning
than flux compactification. For example looking for  scenarios
where the extra dimensions become small, while we see the
expansion in the three space dimensions, in the early stages of
the cosmic evolution, and in particular  during inflation. There
have also been proposals to realize this ``dynamical moduli
fixing'' scenarios within string theory, which usually go under the name
of ``string gas cosmology'' \cite{B-V, B-V-moduli} or their more
recent improvements ``brane gas cosmology'' \cite{Robert}. These
models are based on the fact that in string cosmology we are
dealing with a gas of strings and branes rather than a gas of
point particles described by standard local field theories.

In this paper we examine a different new idea for the dynamical
stabilization of the volume. In our model we restrict ourselves to
Einstein gravity in higher dimensions. In particular we start with
a generalized FRW cosmology plus noncommutativity in the space of
{\it momenta} conjugate to the metric in the extra dimensions and
show that this noncommutativity tames the runaway behavior of the
scale factors of the internal space. The idea of noncommutativity in the context of
cosmology and its relevance to moduli stabilization, has been previously discussed
in the literature.
For example in  \cite{Ardalan} noncommutativity in the extra directions and that
it can help with stabilization of the volume of compactification has been discussed; in
\cite{NC-Cosmology} noncommutativity in the superspace (and not the real physical space)
in the context of four dimensional gravity theory was considered and the noncommutative version
of Wheeler-DeWitt equation discussed. Our model, in a sense, has a combination of the two.
Namely, in our model we deal with a deformed superspace (and not the real physical space)
in the context of a higher dimensional theory. The noncommutativity we consider is
in the momentum part of the mini-super-phase-space and is only in the extra
dimensions.

The rest of the paper is organized as follows. In section 2, we
present a preliminary set up for the model starting with a $D$
dimensional gravity theory and use the $(D-1)+1$ ADM decomposition
to work out the Hamiltonian constraint, through which using
Hamilton-Jacobi dynamics, one can obtain the gravitational
equations of motion. In section 3, we add the other crucial
ingredient of our model, namely turning on the noncommutativity in
the momentum space. We then re-solve the gravitational equations
of motion for the noncommutative case and show that while the
scale factor in the four space-time dimensions has a power-law
growth with respect to the co-moving time, the scale factor in the
extra dimensions exhibits an oscillatory behavior, thus
stabilizing the volume modulus. We end with discussions and
remarks and discuss a possible setting to realize our model within
string theory flux compactifications.

\section{Preliminary set up of the model}
Consider a $D$ dimensional universe defined as
\begin{equation}
{\cal{M}}=R \times S \times M \times N ,
\end{equation}
where $\times$ are wrapped products, $S$ is the spatial part of
our four dimensional space-time and $M,N$ are the internal (extra)
dimensions. We choose $M$ and $N$ to be \emph{compact} $m,n$
dimensional manifolds, hence $D=m+n+4$. We parameterize the line
element as
\begin{equation}\label{metric-ansatz}
ds^2=-e^{2\rho(t)}dt \otimes dt+e^{2a(t)}dx^i \otimes
dx^i+e^{2u(t)}dy^j \otimes dy^j+e^{2v(t)}dz^k \otimes dz^k ,
\end{equation}
where $i=1,2,3$, $j=1,...,m$, $k=1,...,n$ and $a, u$ and $v$
are the scale factors we have associated with $S, M$ and $N$
respectively. Dynamics of the metric is governed by the $D$
dimensional action
\begin{equation}\label{D-action}
{\cal{S}}=\frac{1}{2\kappa_0^2}\int d^Dx \sqrt{g}
R[g]%+{\cal{S}}_{YGH},
\end{equation}
where $R[g]$ is the scalar curvature of the metric
(\ref{metric-ansatz}), $\kappa_0^2$ is the $D$-dimensional
gravitational (Newton) constant.
% and ${\cal{S}}_{YGH}$ is the usual York-Gibbons-Hawking boundary term.
Assuming that all the
fields (the metric components) are only  functions of $t$,
plugging (\ref{metric-ansatz}) into the action (\ref{D-action}) we
obtain
\begin{equation}
{\cal{S}}=\frac{1}{2\kappa^2} \int dtd^3x {\cal{L}},
\end{equation}
where $\kappa$ is the four dimensional Newton constant,
\[
\kappa^2=\frac{\kappa^2_0}{V_M\ V_N},
\]
with $V_M,V_N$ being the volumes of $M,N$ internal manifolds, and
\begin{equation}
{\cal{L}}=\frac{1}{2}e^{-\rho+\rho_0}\left[(3-9)\dot{a}^2+(m-m^2)\dot{u}^2+(n-n^2)\dot{v}^2-
2\left(3m\dot{a}\dot{u}+3n\dot{a}\dot{v}+nm\dot{u}\dot{v}\right)\right].
\end{equation}
Here,  dot represents  differentiation with respect to $t$ and
\begin{eqnarray}
\rho_0(t)=3a(t)+mu(t)+nv(t).
\end{eqnarray}
For the above Lagrangian one may write the corresponding Hamiltonian
as
\begin{equation}\label{Hamiltonian}
{\cal{H}}=\frac{e^{-\rho+\rho_0}}{2+n+m}\left[\alpha p_a^2+\beta
p_u^2+\gamma p_v^2 -p_ap_u-p_ap_v-p_up_v\right],
\end{equation}
where $p_a$, $p_u$ and $p_v$ are the momenta conjugate to $a$, $u$
and $v$ respectively and%
\be\label{alpha-beta-gamma}%
\alpha=\frac{n+m-1}{6}\ ,\qquad \beta=\frac{n+2}{2m}\
,\qquad
\gamma=\frac{m+2}{2n}.%
\ee%
The equations of motion corresponding to this Hamiltonian, noting
that the overall constant factor $\frac{1}{2+n+m}$ has been
dropped, become%
\begin{eqnarray}\label{eom-com-1}
\dot{a}&=&\{a,{\cal{H}}\}_P=e^{-\rho+\rho_0}\left(2\alpha
p_a-p_u-p_v\right),
\quad \dot{p_a}=\{p_a,{\cal{H}}\}_P={\cal{H}}e^{\rho-\rho_0}\{p_a,e^{-\rho+\rho_0}\}_P,\nonumber\\
\dot{u}&=&\{u,{\cal{H}}\}_P=e^{-\rho+\rho_0}\left(2\beta
p_u-p_a-p_v\right),
\quad \dot{p_u}=\{p_u,{\cal{H}}\}_P={\cal{H}}e^{\rho-\rho_0}\{p_u,e^{-\rho+\rho_0}\}_P,
%\nonumber
\\
\dot{v}&=&\{v,{\cal{H}}\}_P=e^{-\rho+\rho_0}\left(2\gamma
p_v-p_a-p_u\right), \quad
\dot{p_v}=\{p_v,{\cal{H}}\}_P={\cal{H}}e^{\rho-\rho_0}\{p_v,e^{-\rho+\rho_0}\}_P.\nonumber
\end{eqnarray}
Since the momentum conjugate to $\rho(t)$ does not appear in
Hamiltonian (\ref{Hamiltonian}), the equation of motion for
$\rho$,
\be\label{const-com}%
{\cal{H}}={\cal{L}}=0,%
\ee%
appears as a constraint which should be imposed on the solutions
of \eqref{eom-com-1}. Therefore, as long as \eqref{const-com} is
satisfied $\rho$ could be taken any function of time. To solve
\eqref{eom-com-1} it is convenient to adopt the {\it harmonic time gauge}%
\begin{eqnarray}
\rho(t)=\rho_0(t)=3a(t)+mu(t)+nv(t). \label{harmonic-time}
\end{eqnarray}
The equations then simplify to
\be\label{eom-com-2}%
\ddot{a}=0,\qquad \ddot{u}=0,\qquad \ddot{v}=0,%
\ee%
whose solutions are of the form%
\be\label{com-sol}%
a(t)=a_1t +b_1, \qquad u(t)=a_2t +b_2,\qquad v(t)=a_3t +b_3,%
\ee%
where $a_i, b_i$ are arbitrary constants. We should now make sure
that the (Hamiltonian) constraint \eqref{const-com}
\be\label{const-sol}%
{\cal M}_{ij} a_ia_j\equiv (3-9)a_1^2+(m-m^2)a_2^2+(n-n^2)a_3^2-
2(3ma_1a_2+3na_1a_3+n m a_2 a_3)=0%
\ee%
can be satisfied. This equation does have solutions for real
$a_i$, because the $3\times 3$ matrix ${\cal M}_{ij}$ has negative
eigenvalues. More concretely it has two positive and one negative
eigenvalues.\footnote{In fact the Hamiltonian \eqref{Hamiltonian}
is not positive definite. This is, however, a well known fact in
gravity e.g. see \cite{Weinberg}. To see this recall that gravity
can be viewed as a gauge theory with the local Lorentz symmetry.}
One may solve \eqref{const-sol} and obtain e.g. $a_3$ as a
function of $a_1$ and $a_2$.

The above result is obtained in the harmonic time gauge. To see
the significance of this result for cosmological purposes it is
better to use the comoving frame. This could be done introducing
the comoving time $\tau$:
\begin{eqnarray}
\frac{d\tau}{dt}=e^{\rho(t)}, \label{eqn1}
\end{eqnarray}
leading to
\be\label{comoving-time}%
\tau=\frac{1}{K}\ e^{Kt},\ \qquad K=3a_1+ma_2+na_3.%
\ee%
In terms of $\tau$, the scale factors $e^a$ becomes
\be%
e^{a(\tau)}\sim \tau^{\frac{a_1}{3a_1+ma_2+na_3}}.%
\ee%
Therefore, for generic values of the initial conditions $a_i$, we
find a power-law behavior in all of the spatial directions and
hence there is no stabilization for the internal dimensions. This
is somewhat expected because the internal scale factors in terms
of the lower dimensional theory, after Kaluza-Klein
compactification, appear as scalar fields with run away potentials
driving the power-law behavior. In the next section we demonstrate
how noncommutativity among the internal conjugate momenta $p_u,
p_v$ can lead to a mechanism which dynamically stabilizes the
volume of the extra dimensions by modifying the run away behavior.

\section{Noncommutative mini-super phase space}

The Hamiltonian \eqref{Hamiltonian} describes a classical (or
quantum) mechanical system on a three dimensional mini-super-space
or a six dimensional mini-super-phase-space.  In the previous
section we used standard canonical phase space with the usual
Poisson brackets. We would now like to modify the canonical
brackets on the mini-super-phase-space. We do this by changing the
Poisson brackets of conjugate momenta to be non-vanishing. Since
$a,u,v$ do not appear in the Hamiltonian explicitly, making them
noncommuting will not modify the equations of motion.

We impose noncommutativity between the internal momenta:%
\be\label{NC-momenta}\begin{split}%
\{P_v,P_u\}_P&=\xi,\cr \{v,u\}_P& =0.
\end{split}\ee%
We'll, however, keep the Hamiltonian to  have the same functional
form in terms of $P$'s as before. Such a noncommutativity can be
motivated by string theory corrections to the Einstein gravity. In
the next section we discuss this point further.

Next we use the standard trick \cite{NC-Lamb} to re-introduce the
canonical variables%
\be\begin{split}
p_u&=P_u+\frac{\xi}{2}v,\\
p_v &=P_v-\frac{\xi}{2}u.%
\end{split}%
\ee%
It is evident that $\{p_v,p_u\}_P=0$. The rest of the analysis
reduces to the standard ones once we rewrite Hamiltonian in terms
of the canonical variables $u,v,
p_u,p_v$%
\begin{eqnarray}\label{NC-H}
%\begin{split}%
{\cal{H}}_{nc}&=&\frac{e^{-\rho+\rho_0}}{2+n+m}\left[\alpha
p_a^2+\beta (p_u-\frac{\xi}{2} v)^2+\right.\gamma
(p_v+\frac{\xi}{2}u)^2\nonumber\\ &-& \left. p_a(p_u-\frac{\xi}{2}
v)-p_a(p_v+\frac{\xi}{2}u)-(p_u-\frac{\xi}{2}v)(p_v
+\frac{\xi}{2}u)\right]
%\end{split}
\end{eqnarray}
where $\alpha$, $\beta$ and $\gamma$ are defined in
\eqref{alpha-beta-gamma}. As we see the Hamiltonian \eqref{NC-H}
is very similarly to that of a charged particle moving in an
external magnetic field proportional to $\xi$ in the $u,v$
directions and the flat potential in the $u$ and $v$ directions is
lifted to a harmonic oscillator potential due to noncommutativity.

The equations of motion can now be easily worked out as
\begin{eqnarray}\label{eom-NC}
\eta\dot{a}&=&e^{-\rho+\rho_0}\left(2\alpha
p_a-p_u-p_v-\frac{\xi}{2}(u-v)\right),\nonumber\\
\eta\dot{p_a}&=&\eta{\cal{H}}_{nc}e^{\rho-\rho_0}\{p_a,e^{-\rho+\rho_0}\}_P,%\nonumber
\\
\eta\dot{u}&=&e^{-\rho+\rho_0}\left(2\beta
p_u-p_a-p_v-\frac{\xi}{2}(2\beta  v+u)\right),\nonumber\\
\eta\dot{p_u}&=&\eta{\cal{H}}_{nc}e^{\rho-\rho_0}\{p_u,e^{-\rho+\rho_0}\}_P+e^{-\rho+\rho_0}
\left(-\gamma \xi
p_v+\frac{\xi}{2}p_a+\frac{\xi}{2}p_u-\frac{\gamma \xi^2}{2}u-
\frac{\xi^2}{4}v\right),%\nonumber
\\
\eta\dot{v}&=&e^{-\rho+\rho_0}\left(2\gamma
p_v-p_a-p_u+\frac{\xi}{2}(2\gamma u +v)\right),\nonumber\\
\eta\dot{p_v}&=&\eta{\cal{H}}_{nc}e^{\rho-\rho_0}\{p_v,e^{-\rho+\rho_0}\}_P+e^{-\rho+\rho_0}
\left(\beta \xi p_u-\frac{\xi}{2}p_a-\frac{\xi}{2}p_v-\frac{\beta
\xi^2}{2}v- \frac{\xi^2}{4}u\right),%\nonumber
\end{eqnarray}
where
\[
\eta=2+m+n=D-2,
\]
 and equation of motion for $\rho$ again appears as the
Hamiltonian constraint
\be\label{const-noncom}%
{\cal{H}}_{nc}=0.%
\ee%
Physically one should expect this, because the Hamiltonian
constraint is the result of time re-parametrization invariance
which remains even when  the noncommutativity is turned on.

 In the harmonic time gauge (\ref{harmonic-time}) the above equations simplify as
\begin{eqnarray}\label{NC-eom}
 \eta\ddot{a}&=&-\xi \left(\dot{u}-\dot{v}\right),\nonumber\\
 \eta\ddot{u}&=&-\xi \left(\dot{u}+2\beta\dot{v}\right),\\
 \eta\ddot{v}&=&\xi \left(\dot{v}+2\gamma\dot{u}\right),\nonumber
\end{eqnarray}
whose solutions are%
\bea\label{NC-sol}%
%\begin{align}
a(t)&=&\frac{b}{\tan\theta}\left[\sqrt{\gamma} \cos(\omega
t+\phi+\theta)+\sqrt{\beta}\cos(\omega t+\phi)\right] +{H_0}
t+C_1,\cr
u(t)&=&b\sqrt{\beta}\sin(\omega t+\phi)+C_2,\\
v(t)&=&-b\sqrt{\gamma}\sin(\omega t+\phi+\theta)+C_3\nonumber,
%\end{align}
\eea%
where $b, C_1, C_2, C_3,H_0$ and $\phi$ are integration constants
and
\begin{eqnarray}
\theta=\arccos\left(\frac{1}{\sqrt{4 \beta \gamma}}\right),\\
\eta\omega=\xi \sqrt{4\beta \gamma-1}=+\xi\tan\theta.
\end{eqnarray}
Since $4\beta\gamma=1+2\left(\frac{m+n+2}{mn}\right)>1$, $\omega$
and $\theta$ are both real valued. It can be easily checked that
the above noncommutative solutions in the $\xi\rightarrow 0$ go
over to the commutative ones (\ref{com-sol}). As we see now $u$
and $v$ show an oscillatory behavior. They have a phase difference
$\theta$ which depends only on the number of the dimensions of $M$
and $N$. For example for $m=n$ case,
\[
\cos\theta=\frac{\# {\rm extra\ dimensions}}{\# {\rm total\
spacetime\ dimensions}}=\frac{n}{n+2}, \qquad
\omega=\frac{1}{n\sqrt{n+1}}\,\,\xi.
\]
For the ten dimensional case, $\cos\theta=3/5$, $\omega=\xi/6$.

We must now impose the Hamiltonian constraint
\eqref{const-noncom}. It is basically an equation for $H_0$.
Plugging the solutions
\eqref{NC-sol} into ${\cal H}_{nc}$ we obtain%
\be\label{H0-xi}%
 H_0^2=\frac{1}{2\times 3mn} b^2\xi^2%
\ee%
As we expect, $H_0$ turns out to be proportional to the only
dimensionful parameter in our problem, $\xi$. Moreover, as one
would physically expect the constants $C_i$ and $\phi$ which could
be absorbed in the scaling of coordinates and choice of origin of
time, does not appear in the final expression for the Hamiltonian
constraint and hence in $H_0$. For the special ten dimensional
case and when $m=n=3$, $H_0=b\xi/3\sqrt{6}$.

Dropping the $C_i$ and $\phi$
the scale factors of the internal dimensions can be written as %
\bea\label{scale-factors}
e^{2u(t)}&=&\exp\left[2b\sqrt{\beta}\sin\omega t\right],\cr
e^{2v(t)}&=&\exp\left[-2b{\sqrt\gamma}\sin(\omega
t+\theta)\right],\\
%\label{3d-scale-factor}%
e^{2a(t)}&=&\exp\left[\frac{2b}{\tan\theta}\left(\sqrt{\gamma}
\cos(\omega t+\theta)+\sqrt{\beta}\cos\omega
t\right)\right]e^{2H_0t},\nonumber
\eea%
where $H_0$ is given by \eqref{H0-xi}. A quick look at the above
equations reveals that as the time $t$ progresses the behavior of
the internal scale factors, the first two equations, is
oscillatory while that of the visible spacetime has a growing
behavior.
\begin{figure}
\begin{tabular}{ccc}
 \epsfig{figure=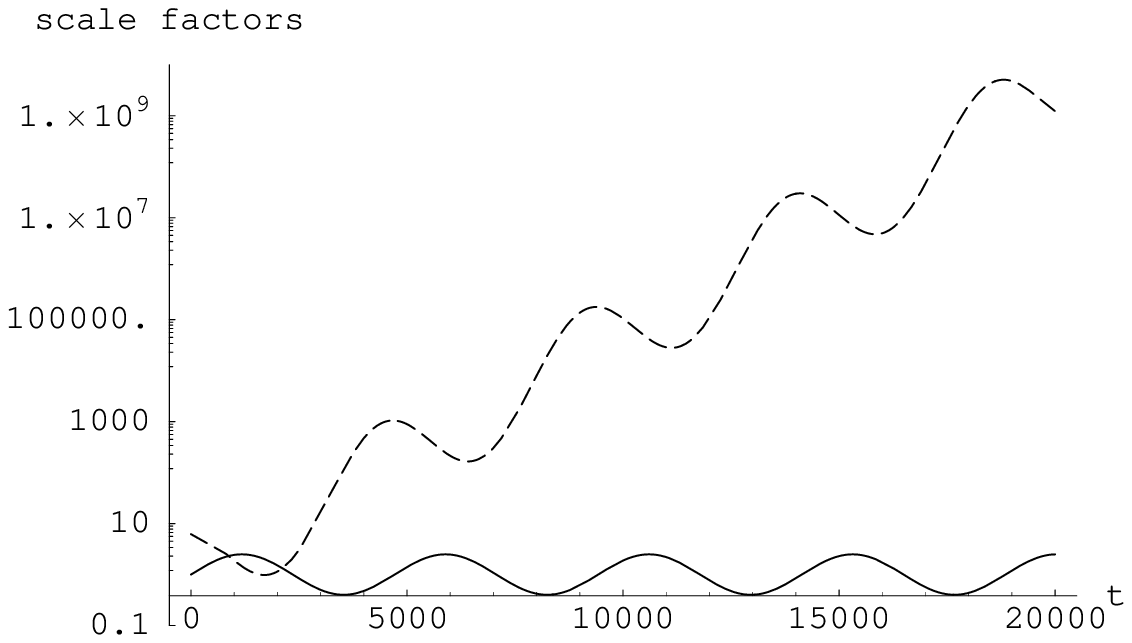,width=8cm}\hspace{1mm}
 \epsfig{figure=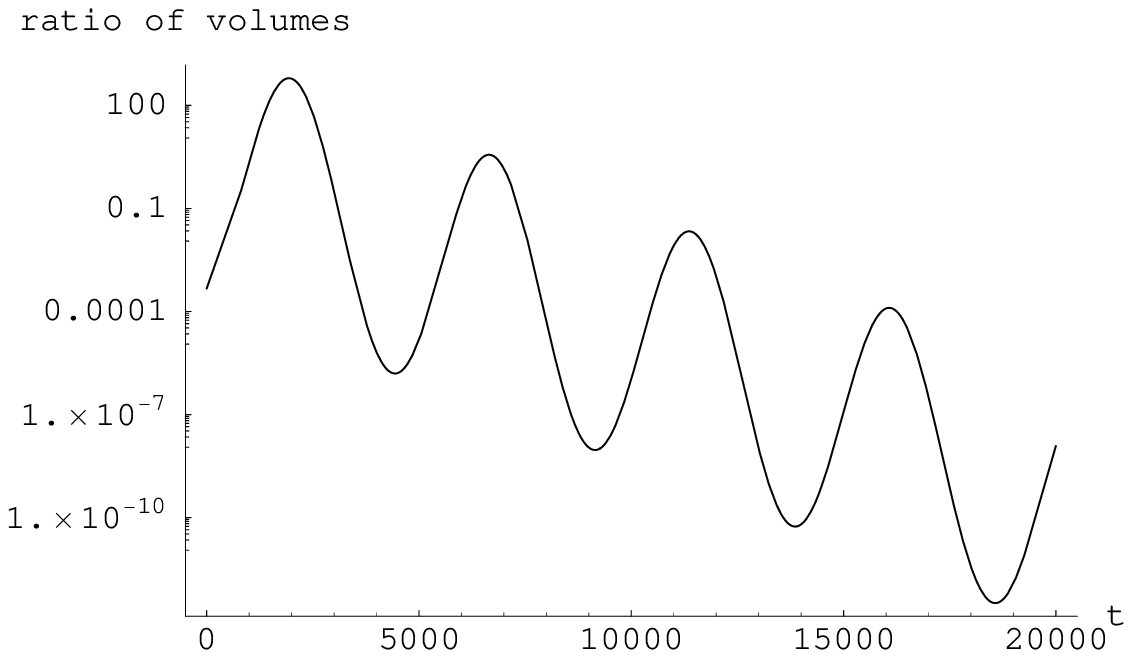, width=8cm}
\end{tabular}
\caption{\footnotesize The left plot shows  the scale factors with
respect to the harmonic time $t$ measured in units of $H_0^{-1}$.
The harmonic time coordinate $t$ is essentially logarithm of the
comoving time $\tau$ (for large $t$). These figures are drawn to
demonstrate the oscillatory behavior. The solid line represents
the scale factor for the extra dimensions using equation
(\ref{scale-factors}). The dashed line represents the scale factor
of the ordinary universe. The plot on the right shows the ratio of
the volume of the internal space to that of the ordinary space.
These figures are drawn for $n=m=3$, $b=1$ and $\xi\sim H_0$. }
\label{fig1}
\end{figure}

To gain a better understanding of this behavior it is useful to
work in the more standard comoving FRW frame. As discussed in the
previous section for this we need to redefine the time coordinate.
To simplify the computations let
us consider the special case which is of particular interest to string theory,%
\[
m=n=3,
\]
corresponding to a ten dimensional space-time. For this case
(\ref{harmonic-time}) becomes,%
\bea\label{NCcomoving-time} \rho=\frac{b\xi}{\sqrt{6}}
t+\frac{3b}{\sqrt{6}} \cos\left(\frac{\xi}{6}
t+\frac{\theta}{2}\right)
\eea%
and the co-moving time $\tau$ is given by $d\tau=e^\rho dt$. The
form of the scale factors and the volume ratio with respect to the
co-moving time $\tau$ are drawn in the Figures. For the present
epoch, that is for $t\gg \frac{1}{ \xi}$ in
(\ref{NCcomoving-time}), the first term has the dominant
contribution and one can write $ \tau=\frac{1}{K}\ e^{Kt},\quad
K=3H_0=b\xi/\sqrt{6}, $ similarly to the commutative case
(\ref{comoving-time}).

\begin{figure}
\begin{tabular}{ccc}
\epsfig{figure=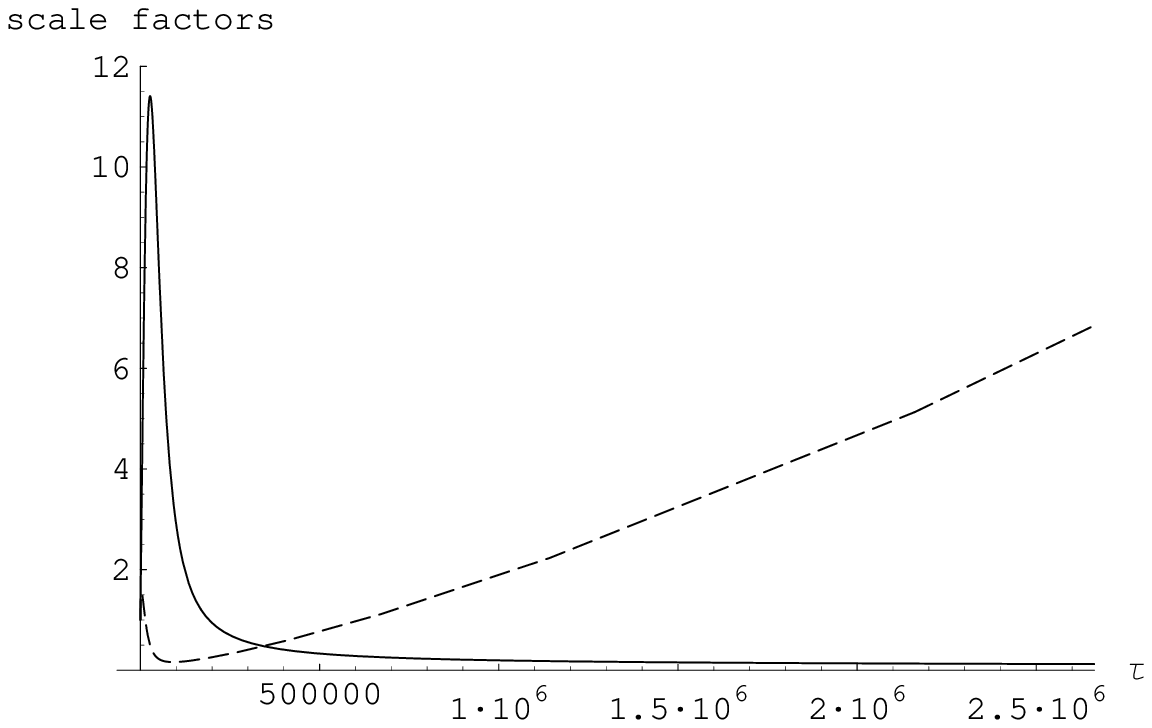, width=8cm}
\hspace{0.1cm}
\epsfig{figure=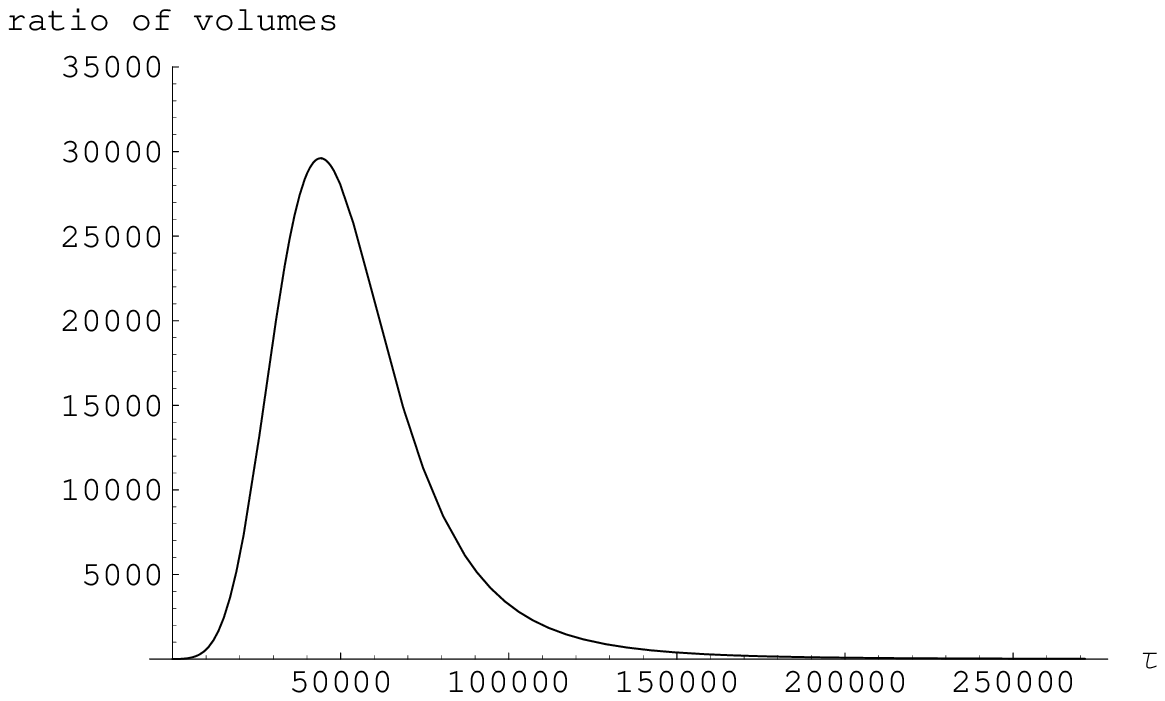, width=8cm}
\end{tabular}
\caption{\footnotesize The left plot shows  the scale factors with
respect to the co-moving time $\tau$ measured in units of
$H_0^{-1}$. The solid line represents the scale factor for the
extra dimensions using  equation (\ref{scale-factors}). The dashed
line represents the scale factor of the ordinary universe. The
plot on the right shows the ratio of the volume of the internal
space to that of the ordinary space. These figures are drawn for
$n=m=3$, $b=1$ and $\xi\sim H_0$. Note that the expected
oscillatory behavior, not visible here, appears in later times.
The same is true for the power law behavior shown by the scale
factor of the ordinary universe, the dashed line.} \label{fig2}
\end{figure}

\section{Discussion and concluding remarks}

Here we discussed a simple  model for stabilizing compactification
volume in a higher dimensional gravity theory. Starting  from a
$D=m+n+4$ dimensional spacetime with all the spatial directions
compactified, say on a torus of Planckian size, we showed that the
assumption that the momenta conjugate to scale factors of the
internal degrees are noncommuting plus the usual Einstein
equations, lead to the very interesting result that the scale
factor corresponding to three visible spatial dimensions obey a
power law growth (see the figures), while the internal scale
factors just oscillate. The solutions to our model involves two
time scales, one is the noncommutativity scale $\xi$ and the other
is $H_0$, whose ratio $b$ is an integration constant (an initial
condition).  The natural choice is of course $b\sim 1$ or $\xi\sim
H_0$ in which case as is seen from the figures, the ratios of the
internal space to the visible space volumes drops by a power law
in the comoving cosmological time $\tau$.

In short, introduction of  noncommutativity among the momenta in
the internal part of mini-super-phase-space modifies the
gravitational dynamics in such a way that the average of the
volume of the internal compact directions is stable under the time
evolution of the system. In a sense, as is also seen from
(\ref{NC-H}), introduction of noncommutativity among the momenta
for the internal scale factors is like the addition of a harmonic
oscillator type potential for the internal directions. In other
wording, the scale factors of the internal dimensions appear as
scalars in the $3+1$ dimensional theory and the noncommutativity
we have added removes the run away
 behavior of these scalars. This is essentially the same
phenomenon that particles in an external magnetic field (the
Landau problem) have effectively noncommuting momenta.

In our model we have assumed noncommuting momenta
\eqref{NC-momenta} which needs to be motivated. As mentioned
earlier and is well-known  noncommuting momenta arises when we
have a charged particle in a magnetic field and the Landau
problem. However, here we have a gravitational system and it is
not clear what this ``magnetic field'' can be. Although we do not
have a robust setup in a gravitational context which leads to
\eqref{NC-momenta}, one might hope that turning on fluxes in the
internal compact space in a string theory setting and within flux
compactification models, can lead to noncommuting momenta in the
internal part of the mini-super-phase-space. (Recall that flux
compactifications was proposed as a way to stabilize some of the
compactification moduli.) This is in the same spirit as the
noncommutativity appearing on a D-brane in a Kalb-Ramond two from
flux background \cite{NCDbrane}. Here for concreteness and just to
demonstrate the idea we limited ourselves to a constant
noncommutativity among the momenta. We, however, believe that this
effect is not particular to this special case and is a generic
behavior of any similar noncommutativity among momenta. It is
desirable to check this explicitly and if indeed our model can be
fit within string theory settings.

In this work we considered a higher dimensional pure gravity with
no other additional fields. It is of course very interesting to
examine the noncommutativity idea in an inflationary setup and
check if our volume stabilization method still remains effective
in an inflation model.

\vspace{3mm}\noindent\\
{\bf Acknowledgements}\vspace{1mm}\noindent\\
The authors would like to thank Robert Brandenberger for useful
comments on the manuscript and Hassan Firouzjahi for fruitful
discussions.

\end{document}